\documentclass[12pt]{article}
\usepackage{amsbsy}
\usepackage{amsfonts}
\usepackage{amsmath}
\usepackage{amssymb}
\usepackage{apacite}
\usepackage{setspace}
\usepackage{graphicx}
\usepackage{bm,multicol}
\usepackage[english]{babel}
\usepackage{natbib}
\usepackage{hyperref}
\usepackage[T1]{fontenc}
\usepackage[utf8]{inputenc}
\usepackage{authblk}
\usepackage{xcolor}
\hypersetup{colorlinks=false,linkcolor=blue,urlcolor=blue}

\newcommand{\newc}{\newcommand}
\newc{\N}{\mbox{N}}
\newc{\1}{\bf{1}}

\begin{document}
\title{Comment on Garc\'{i}a-Donato et al. (2025) ``Model uncertainty and missing data: An objective Bayesian perspective''}

\author{Joris Mulder}

\thispagestyle{empty} \maketitle
%
%
%

\cite{garcia2025model} present a methodology for handling missing data in a model selection problem using an objective Bayesian approach. The current comment discusses an alternative, existing objective Bayesian method for this problem. First, rather than using the $g$ prior, O'Hagan's fractional Bayes factor \citep[FBF;][]{OHagan:1995} is utilized based on a minimal fraction\footnote{Note that the implied fractional prior has a similar covariance structure as the $g$ prior \citep[e.g., see][]{Mulder:2014b}.}. Second, and more importantly due to the focus on missing data, Rubin’s rules for multiple imputation can directly be used as the fractional Bayes factor can be written as a Savage-Dickey density ratio for a variable selection problem. This attractive property of a Savage-Dickey density ratio was shown by \cite{Hoijtink:2018b}. The use of (adjusted) fractional Bayes factors for a testing problem of a set of predefined equality and/or one-sided constrained hypotheses in the case of missing data was discussed by \cite{mulder2022bayesian}. The current comment derives the methodology for a variable selection problem (which is a special case of the above testing problem). Moreover, its implied behavior is illustrated in a numerical experiment, showing competitive results as the method of \cite{garcia2025model}. Throughout this comment, the same notation is used as \cite{garcia2025model}.

The FBF of a model $\bm\gamma\in\Gamma$ against the full model where $\bm\gamma_{full}=\textbf{1}$ can be written as a Savage-Dickey density ratio \citep[e.g.,][]{mulder2022bayesian,Dickey:1971}
\begin{equation}
B^F_{\bm\gamma,\bm\gamma_{full}}(b) \equiv \frac{m_{\bm\gamma}(\textbf{d})/m_{\bm\gamma}(\textbf{d}^b)}
{m_{\bm\gamma_{full}}(\textbf{d})/m_{\bm\gamma_{full}}(\textbf{d}^b)}
=
\frac{\pi_{\bm\gamma_{full}}(\bm\beta_{-\bm\gamma}=\textbf{0}|\textbf{d})}{\pi_{\bm\gamma_{full}}(\bm\beta_{-\bm\gamma}=\textbf{0}|\textbf{d}^b)},
\label{eqSD}
\end{equation}
where $\bm\beta_{-\bm\gamma}$ denotes the vector of coefficients under the full model which are excluded in model $\bm\gamma$, $m_{\bm\gamma}(\textbf{d}^b)=\iint f_{\bm\gamma}(\textbf{d}|\bm\beta_{\bm\gamma},\bm\alpha)^b\pi^N(\bm\beta_{\bm\gamma},\bm\alpha)d\bm\beta_{\bm\gamma}d\bm\alpha$ denotes the marginal likelihood of model $\bm\gamma$ when raising the likelihood to a fraction $b$, and the numerator (denominator) on the right hand side denotes the marginal posterior \citep[marginal fractional prior; e.g., see][]{Gilks:1995} under the full model evaluated at the null values of the excluded parameters, i.e., $\bm\beta_{-\bm\gamma}=\textbf{0}$. Thus, the FBF can be written as a ratio of a posterior and a fractional prior quantity under the full model.

Hence, we can directly apply Rubin's rules for multiple imputation under the full model. For the posterior, this implies
\begin{eqnarray}
\nonumber\pi_{\bm\gamma_{full}}(\bm\beta_{-\bm\gamma}=\textbf{0}|\textbf{d}_{(0)})&=&\int \pi_{\bm\gamma_{full}}(\bm\beta_{-\bm\gamma}=\textbf{0}|\textbf{d}_{(0)},\textbf{d}_{(1)})m_{\bm\gamma_{full}}(\textbf{d}_{(1)}|\textbf{d}_{(0)})d\textbf{d}_{(1)}\\
\label{postMI}&=&\text{AVE}[\pi_{\bm\gamma_{full}}(\bm\beta_{-\bm\gamma}=\textbf{0}|\textbf{d}_{(0)},\textbf{d}_{(1)})]
\end{eqnarray}
where $\text{AVE}[~\cdot ~]$ refers to the average based on repeated imputations drawn from the posterior predictive distribution of missing data given the observed data under the full model, $m_{\bm\gamma_{full}}(\textbf{d}_{(1)}|\textbf{d}_{(0)})$ \citep[e.g.,][]{Rubin:1996}. For the fractional prior, we write
\begin{eqnarray}
\nonumber\pi_{\bm\gamma_{full}}(\bm\beta_{-\bm\gamma}=\textbf{0}|\textbf{d}_{(0)}^b)&\equiv&\int \pi_{\bm\gamma_{full}}(\bm\beta_{-\bm\gamma}=\textbf{0}|(\textbf{d}_{(0)},\textbf{d}_{(1)})^b)m_{\bm\gamma_{full}}(\textbf{d}_{(1)}|\textbf{d}_{(0)})d\textbf{d}_{(1)}\\
\label{priorMI}&=&\text{AVE}[\pi_{\bm\gamma_{full}}(\bm\beta_{-\bm\gamma}=\textbf{0}|(\textbf{d}_{(0)},\textbf{d}_{(1)})^b)].
\end{eqnarray}
Hence, similar as for the posterior quantity, the posterior predictive distribution based on all observed data, $\textbf{d}_{(0)}$, is used to compute the fractional prior quantity. Note that if a minimal fraction of the observed data would have been used in the predictive distribution, the imputed missing data would be unrealistically heterogeneous. Moreover, by taking a fraction of the observed data and the missing data, i.e., $\pi_{\bm\gamma_{full}}(\bm\beta_{-\bm\gamma}=\textbf{0}|(\textbf{d}_{(0)},\textbf{d}_{(1)})^b)$, the fractional prior is again based on a fraction of the information in the observed data, similar as in the fractional Bayes factor. This construction also allows us to compute the posterior and fractional prior quantities using the same imputed data. Moreover, as the marginal posterior and marginal fractional prior of $\bm\beta_{-\bm\gamma}$ both have multivariate Student $t$ distributions \citep[e.g.,][]{mulder2022bayesian}, computation is straightforward. The R package \texttt{BFpack} \citep{mulder2021bfpack} computes these quantities for fractional Bayes factors for any equality/one-sided constrained model, which also includes the model $\bm\gamma$. Finally note that by separately computing the numerator and denominator in \eqref{eqSD} for the observed data, the fractional Bayes factor is still coherent via this construction \citep[see also][]{OHagan:1997}.

\begin{figure}[t]
\begin{center}
\includegraphics[width=\textwidth]{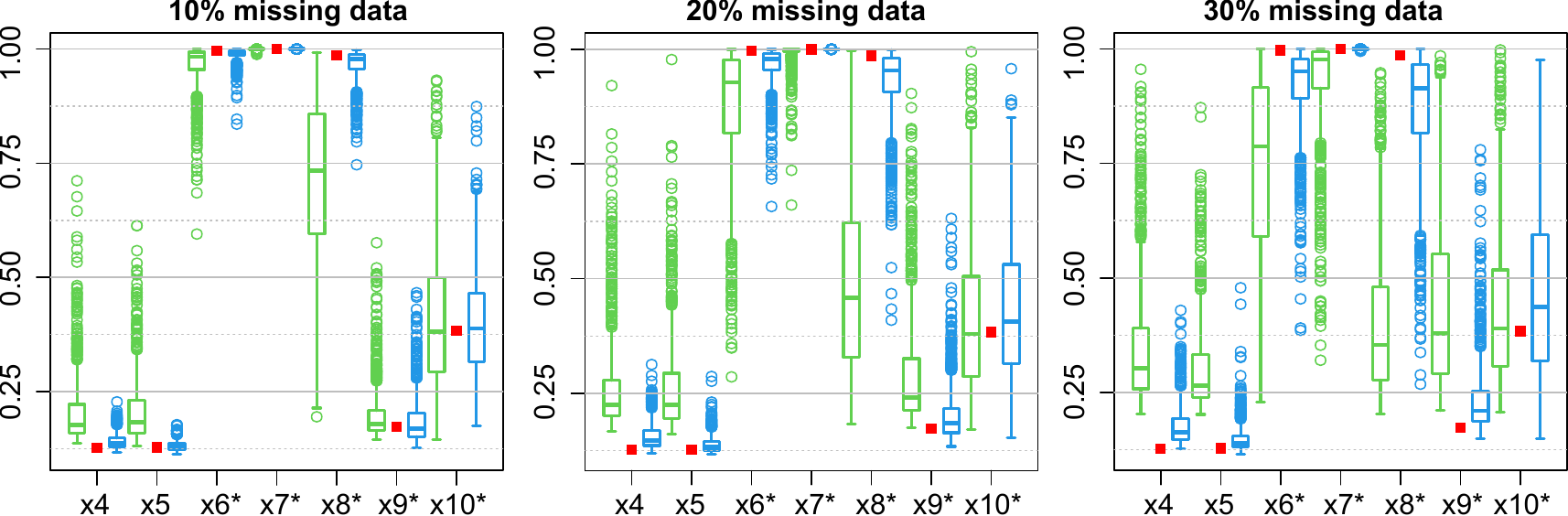}
\caption{Boxplots of the inclusion probabilities for each variable using FBFs after list-wise deletion (green) and imputed FBFs via Savage-Dickey ratios (blue) when considering a proportion of 0.10 (left), 0.20 (middle), or 0.30 (right) of missing values per variable, for the Ozone dataset. The inclusion probabilities based on the oracle FBF using the full dataset are depicted as red squares. The variable names with `$*$' contained missing observations.}
\label{sim_fig_miss}
\end{center} 
\end{figure}

I end this comment by illustrating the implied selection behavior of this method for Experiment 2 of \cite{garcia2025model} using the \texttt{Ozone35} dataset with 7 potential predictors and the same missing data mechanisms. Inclusion probabilities were obtained using FBFs after list-wise deletion and when using FBFs using the above methodology for handling missing data. The R package \texttt{BFpack} was used for computing the posterior probabilities for all $7^2=128$ possible models from which the inclusion probabilities can be computed\footnote{The R code can be found here: \href{https://github.com/jomulder/missing-data-using-BFpack-with-FBFs/tree/main}{http://github.com/jomulder/missing-data-BFpack-FBFs}.}. Figure \ref{sim_fig_miss} shows the boxplots of the inclusion probabilities for all predictors based on proportions of missing values for the variables $x_6$ to $x_{10}$ of 10\%, 20\%, or 30\% when using list-wise deletion (green plots) and the method discussed above (blue plots). Similar as Garc\'{i}a-Donato et al.'s method, the proposed method is clearly superior over list-wise deletion by better preserving the evidence. Moreover, the resulting inclusion probabilities using this method are very similar as compared to Garc\'{i}a-Donato et al.'s method (comparing Figure \ref{sim_fig_miss} here with Figure 1 of Garc\'{i}a-Donato et al.).

\bibliographystyle{apacite}
\bibliography{refs_mulder}

\begin{thebibliography}{}

\bibitem [\protect \citeauthoryear {%
Dickey%
}{%
Dickey%
}{%
{\protect \APACyear {1971}}%
}]{%
Dickey:1971}
\APACinsertmetastar {%
Dickey:1971}%
\begin{APACrefauthors}%
Dickey, J.%
\end{APACrefauthors}%
\unskip\
\newblock
\APACrefYearMonthDay{1971}{}{}.
\newblock
{\BBOQ}\APACrefatitle {The Weighted Likelihood Ratio, Linear Hypotheses on
  Normal Location Parameters} {The weighted likelihood ratio, linear hypotheses
  on normal location parameters}.{\BBCQ}
\newblock
\APACjournalVolNumPages{The Annals of Statistics}{42}{1}{204--223}.
\PrintBackRefs{\CurrentBib}

\bibitem [\protect \citeauthoryear {%
Garc{\'\i}a-Donato%
, Castellanos%
, Cabras%
, Quir{\'o}s%
\BCBL {}\ \BBA {} Forte%
}{%
Garc{\'\i}a-Donato%
\ \protect \BOthers {.}}{%
{\protect \APACyear {2025}}%
}]{%
garcia2025model}
\APACinsertmetastar {%
garcia2025model}%
\begin{APACrefauthors}%
Garc{\'\i}a-Donato, G.%
, Castellanos, M\BPBI E.%
, Cabras, S.%
, Quir{\'o}s, A.%
\BCBL {}\ \BBA {} Forte, A.%
\end{APACrefauthors}%
\unskip\
\newblock
\APACrefYearMonthDay{2025}{}{}.
\newblock
{\BBOQ}\APACrefatitle {Model Uncertainty and Missing Data: An Objective
  {B}ayesian Perspective} {Model uncertainty and missing data: An objective
  {B}ayesian perspective}.{\BBCQ}
\newblock
\APACjournalVolNumPages{{B}ayesian Analysis}{1}{1}{1--26}.
\PrintBackRefs{\CurrentBib}

\bibitem [\protect \citeauthoryear {%
Gilks%
}{%
Gilks%
}{%
{\protect \APACyear {1995}}%
}]{%
Gilks:1995}
\APACinsertmetastar {%
Gilks:1995}%
\begin{APACrefauthors}%
Gilks, W\BPBI R.%
\end{APACrefauthors}%
\unskip\
\newblock
\APACrefYearMonthDay{1995}{}{}.
\newblock
{\BBOQ}\APACrefatitle {Discussion to fractional {B}ayes factors for model
  comparison (by {O}'{H}agan)} {Discussion to fractional {B}ayes factors for
  model comparison (by {O}'{H}agan)}.{\BBCQ}
\newblock
\APACjournalVolNumPages{Journal of the Royal Statistical Society Series
  B}{56}{}{118--120}.
\PrintBackRefs{\CurrentBib}

\bibitem [\protect \citeauthoryear {%
Hoijtink%
, Gu%
, Mulder%
\BCBL {}\ \BBA {} Rosseel%
}{%
Hoijtink%
\ \protect \BOthers {.}}{%
{\protect \APACyear {2018}}%
}]{%
Hoijtink:2018b}
\APACinsertmetastar {%
Hoijtink:2018b}%
\begin{APACrefauthors}%
Hoijtink, H.%
, Gu, X.%
, Mulder, J.%
\BCBL {}\ \BBA {} Rosseel, Y.%
\end{APACrefauthors}%
\unskip\
\newblock
\APACrefYearMonthDay{2018}{}{}.
\newblock
{\BBOQ}\APACrefatitle {Computing {B}ayes Factors From Data With Missing Values}
  {Computing {B}ayes factors from data with missing values}.{\BBCQ}
\newblock
\APACjournalVolNumPages{Psychological Methods}{24}{2}{253--268}.
\newblock
\begin{APACrefDOI} \doi{10.1037/met0000187} \end{APACrefDOI}
\PrintBackRefs{\CurrentBib}

\bibitem [\protect \citeauthoryear {%
Mulder%
}{%
Mulder%
}{%
{\protect \APACyear {2014}}%
}]{%
Mulder:2014b}
\APACinsertmetastar {%
Mulder:2014b}%
\begin{APACrefauthors}%
Mulder, J.%
\end{APACrefauthors}%
\unskip\
\newblock
\APACrefYearMonthDay{2014}{}{}.
\newblock
{\BBOQ}\APACrefatitle {Prior Adjusted Default {B}ayes Factors for Testing
  (In)equality Constrained Hypotheses} {Prior adjusted default {B}ayes factors
  for testing (in)equality constrained hypotheses}.{\BBCQ}
\newblock
\APACjournalVolNumPages{Computational Statistics and Data
  Analysis}{71}{}{448--463}.
\newblock
\begin{APACrefDOI} \doi{10.1016/j.csda.2013.07.017} \end{APACrefDOI}
\PrintBackRefs{\CurrentBib}

\bibitem [\protect \citeauthoryear {%
Mulder%
\ \BBA {} Gu%
}{%
Mulder%
\ \BBA {} Gu%
}{%
{\protect \APACyear {2022}}%
}]{%
mulder2022bayesian}
\APACinsertmetastar {%
mulder2022bayesian}%
\begin{APACrefauthors}%
Mulder, J.%
\BCBT {}\ \BBA {} Gu, X.%
\end{APACrefauthors}%
\unskip\
\newblock
\APACrefYearMonthDay{2022}{}{}.
\newblock
{\BBOQ}\APACrefatitle {{B}ayesian testing of scientific expectations under
  multivariate normal linear models} {{B}ayesian testing of scientific
  expectations under multivariate normal linear models}.{\BBCQ}
\newblock
\APACjournalVolNumPages{Multivariate Behavioral Research}{57}{5}{767--783}.
\newblock
\begin{APACrefDOI} \doi{10.1080/00273171.2021.1904809} \end{APACrefDOI}
\PrintBackRefs{\CurrentBib}

\bibitem [\protect \citeauthoryear {%
Mulder%
\ \protect \BOthers {.}}{%
Mulder%
\ \protect \BOthers {.}}{%
{\protect \APACyear {2021}}%
}]{%
mulder2021bfpack}
\APACinsertmetastar {%
mulder2021bfpack}%
\begin{APACrefauthors}%
Mulder, J.%
, Williams, D\BPBI R.%
, Gu, X.%
, Tomarken, A.%
, B{\"o}ing-Messing, F.%
, Olsson-Collentine, A.%
\BDBL {}others%
\end{APACrefauthors}%
\unskip\
\newblock
\APACrefYearMonthDay{2021}{}{}.
\newblock
{\BBOQ}\APACrefatitle {{BF}pack: Flexible {B}ayes factor testing of scientific
  theories in R} {{BF}pack: Flexible {B}ayes factor testing of scientific
  theories in r}.{\BBCQ}
\newblock
\APACjournalVolNumPages{Journal of Statistical Software}{100}{}{1--63}.
\PrintBackRefs{\CurrentBib}

\bibitem [\protect \citeauthoryear {%
O'Hagan%
}{%
O'Hagan%
}{%
{\protect \APACyear {1995}}%
}]{%
OHagan:1995}
\APACinsertmetastar {%
OHagan:1995}%
\begin{APACrefauthors}%
O'Hagan, A.%
\end{APACrefauthors}%
\unskip\
\newblock
\APACrefYearMonthDay{1995}{}{}.
\newblock
{\BBOQ}\APACrefatitle {Fractional {B}ayes Factors for Model Comparison (with
  discussion)} {Fractional {B}ayes factors for model comparison (with
  discussion)}.{\BBCQ}
\newblock
\APACjournalVolNumPages{Journal of the Royal Statistical Society
  B}{57}{1}{99--138}.
\newblock
\begin{APACrefDOI} \doi{10.1111/j.2517-6161.1995.tb02017.x} \end{APACrefDOI}
\PrintBackRefs{\CurrentBib}

\bibitem [\protect \citeauthoryear {%
O'Hagan%
}{%
O'Hagan%
}{%
{\protect \APACyear {1997}}%
}]{%
OHagan:1997}
\APACinsertmetastar {%
OHagan:1997}%
\begin{APACrefauthors}%
O'Hagan, A.%
\end{APACrefauthors}%
\unskip\
\newblock
\APACrefYearMonthDay{1997}{}{}.
\newblock
{\BBOQ}\APACrefatitle {Properties of intrinsic and fractional {B}ayes factors}
  {Properties of intrinsic and fractional {B}ayes factors}.{\BBCQ}
\newblock
\APACjournalVolNumPages{Test}{6}{1}{101--118}.
\newblock
\begin{APACrefDOI} \doi{10.1007/BF02564428} \end{APACrefDOI}
\PrintBackRefs{\CurrentBib}

\bibitem [\protect \citeauthoryear {%
Rubin%
}{%
Rubin%
}{%
{\protect \APACyear {1996}}%
}]{%
Rubin:1996}
\APACinsertmetastar {%
Rubin:1996}%
\begin{APACrefauthors}%
Rubin, D\BPBI B.%
\end{APACrefauthors}%
\unskip\
\newblock
\APACrefYearMonthDay{1996}{}{}.
\newblock
{\BBOQ}\APACrefatitle {Multiple Imputation After 18+ Years} {Multiple
  imputation after 18+ years}.{\BBCQ}
\newblock
\APACjournalVolNumPages{Journal of the American statistical
  Association}{91}{434}{473--489}.
\newblock
\begin{APACrefDOI} \doi{10.2307/2291635} \end{APACrefDOI}
\PrintBackRefs{\CurrentBib}

\end{thebibliography}

%
%

\end{document}